# Observation of transverse Thomson effect

Atsushi Takahagi[1,*], Takamasa Hirai[2], Abdulkareem Alasli[1], Sang Jun Park[2], Hosei Nagano[1], and Ken-ichi Uchida[2,3,*]

[1] Department of Mechanical Systems Engineering, Nagoya University, Nagoya 464-8601, Japan

[2] National Institute for Materials Science, Tsukuba 305-0047, Japan

[3] Department of Advanced Materials Science, Graduate School of Frontier Sciences, The University of Tokyo, Kashiwa 277-8561, Japan

[*] Authors to whom correspondence should be addressed:
takahagi.atsushi.u3@s.mail.nagoya-u.ac.jp and UCHIDA.Kenichi@nims.go.jp

**The thermoelectric Thomson effect, predicted in the 1850s by William Thomson, produces volumetric heating/cooling in a conductor due to the concerted action of the Seebeck and Peltier effects. Recently, transverse thermoelectrics studies on the Nernst and Ettingshausen effects have progressed rapidly to enable versatile thermal management technologies and to explore topological transport properties. However, a transverse Thomson effect, arising from the concerted action of the Nernst and Ettingshausen effects, has not yet been observed. Here, we report the observation of the transverse Thomson effect in a conductor. We observed volumetric heating/cooling in a semimetallic $Bi_{88}Sb_{12}$ alloy induced by a charge current, temperature gradient, and magnetic field applied orthogonally to each other using thermoelectric imaging techniques. We found that the heating/cooling can be switched by the field direction. Our experiments and analyses reveal the essential difference between the conventional and transverse Thomson effects; the former depends sorely on the temperature derivative of the Seebeck coefficient, while the latter depends not only on the temperature derivative of the Nernst coefficient but also on its magnitude. The observation of the transverse Thomson effect fills a missing piece in the history of thermoelectrics and provides a new principle for active thermal management technologies.**

The Thomson effect generates volumetric heating/cooling when a charge current and temperature gradient $\nabla T$ are applied in the same direction within a conductor[1,2] (Fig. 1c). The heat production rate per unit volume due to the Thomson effect is given by

$$\dot{q}_{TE} = -\tau_{TE}\mathbf{j}_c \cdot \nabla T \qquad (1)$$

where $\tau_{TE}$ and $\mathbf{j}_c$ are the Thomson coefficient and charge current density, respectively. This effect originates from the longitudinal thermoelectric effects that interconvert charge and heat currents in parallel directions, that is, the Seebeck[3] and Peltier[4] effects (Fig. 1a,b). In the presence of $\nabla T$, the Peltier heat current generated by $\mathbf{j}_c$ varies with position when the longitudinal thermoelectric coefficient of a conductor has finite temperature $T$ dependence. Consequently, Thomson heat release or absorption occurs due to the finite divergence of the heat current in the bulk of the conductor. Through the Onsager reciprocal relation between the Seebeck and Peltier effects[5,6], $\tau_{\text{TE}}$ is determined by the $T$ derivative of the Seebeck coefficient $S_{\text{SE}}$ as follows:

$$\tau_{\text{TE}} = T\frac{dS_{\text{SE}}}{dT} \tag{2}$$

This is known as the first Thomson (or Kelvin) relation[1,2]. While the Seebeck coefficient is regarded as a constant parameter in linear-response thermoelectrics, its $T$ dependence is essential for the Thomson effect. Therefore, the Thomson effect is classified as a higher-order longitudinal thermoelectric phenomenon.

Transverse thermoelectric effects that interconvert charge and heat currents in perpendicular directions have attracted attention for their relevance to fundamental physics and thermal management applications[7-14]. The Nernst[15] (Ettingshausen[16]) effect is a representative transverse thermoelectric effect that generates a charge (heat) current in the direction of the cross-product of an applied heat (charge) current and magnetic field in a conductor (Fig. 1d,e). Although the performance of longitudinal thermoelectric modules reduces due to the presence of multiple junctions in complex Π-shaped thermopile structures, transverse thermoelectric modules based on the Nernst and Ettingshausen effects can be constructed without such junctions, preserving the energy conversion efficiency of the material[17-19]. Owing to these advantages, materials with high transverse thermoelectric conversion performances are actively being investigated for practical applications. As a part of these efforts, a giant Nernst effect was discovered in Dirac and Weyl semimetals[9,12].

These thermoelectric effects, depicted in Fig. 1a-e, were discovered in the 19th century[1-4,15,16] and have been systematically studied for a long time. However, higher-order transverse thermoelectric effects in conductors remain unobserved and the transverse Thomson effect (TTE) is one of such phenomena. Phenomenologically, TTE is expected to occur when a charge current, temperature gradient, and magnetic field are applied orthogonally to each other in a conductor. In the presence of transverse $\nabla T$, the transverse heat current generated by the Ettingshausen effect varies with position when the transverse thermoelectric coefficient of the conductor has finite $T$ dependence, producing volumetric heat release or absorption (Fig. 1f). However, the experimental verification of TTE has not progressed due to challenges in distinguishing it from other thermal effects. Here, we report the observation of TTE in a conductor. We achieved this by developing a thermoelectric imaging technique that isolates TTE signals from temperature modulations caused by the Peltier and Ettingshausen effects, as detailed below.

Despite their phenomenological similarity, TTE fundamentally differs from the longitudinal Thomson effect owing to its transverse nature. The heat production rate per unit volume due to TTE is expressed as

$$\dot{q}_{\text{TTE}} \equiv \tau_{\text{TTE}}\left(\frac{\mathbf{H}}{|\mathbf{H}|} \times \mathbf{j_c}\right) \cdot \nabla T \tag{3}$$

$$\tau_{\text{TTE}} = T\frac{dS_{\text{NE}}}{dT} + 2S_{\text{NE}} \tag{4}$$

where $S_{NE}$ is the Nernst coefficient and the Onsager reciprocal relation between the Nernst and Ettingshausen effects is assumed (Methods)[20]. Importantly, $\tau_{TTE}$ is determined not only by the $T$ derivative of the Nernst coefficient (first term on the right-hand side of equation (4)) but also by its magnitude (second term on the right-hand side of equation (4)), while longitudinal $\tau_{TE}$ is determined sorely by the $T$ derivative of the Seebeck coefficient (equation (2)); the first Thomson relation does not apply to TTE. When heat release or absorption owing to TTE is measured, $\tau_{TTE}$ becomes the experimentally observable quantity rather than $T(dS_{NE}/dT)$.

A material with large $\tau_{TTE}$, that is, both a large magnitude and $T$ dependence of $S_{NE}$, should be selected to observe TTE. Here, we focus on the semimetallic $Bi_{88}Sb_{12}$ alloy, which exhibits a strong magnetic-field-dependent ordinary Nernst effect around room temperature[21]. A polycrystalline $Bi_{88}Sb_{12}$ alloy was synthesized using the spark plasma sintering method under the same conditions as those described in ref.[14,22]. Figure 2a presents the $T$ dependence of $S_{NE}$ of our $Bi_{88}Sb_{12}$ slab measured by the method in ref.[23,24]. To obtain the pure $S_{NE}$ values free from parasitic thermoelectric effects, we paid close attention to the thermal boundary conditions during the measurements (Methods). We confirmed that the $Bi_{88}Sb_{12}$ slab exhibits a large magnitude and finite $T$ dependence of $S_{NE}$ around room temperature. Figure 2b shows the $H$ dependence of $\tau_{TTE}$ and $T(dS_{NE}/dT)$ for $Bi_{88}Sb_{12}$ at 320 K, calculated using the measured $S_{NE}$ values. The sign of $T(dS_{NE}/dT)$ is positive for all $H$ values, whereas that of $\tau_{TTE}$ switches from positive to negative as $H$ increases because of the competition between the first and second terms on the right-hand side of equation (4).

Figure 2c illustrates the TTE measurement system. To clarify the temperature distribution induced by TTE, we adopted a thermoelectric imaging technique based on lock-in thermography[25-28]. A rectangular $Bi_{88}Sb_{12}$ slab sample was fixed to two plates with temperature control modules (Methods). Uniform $\nabla T$ was applied to the sample along the $y$ direction while maintaining the average temperature at 320 K, where the temperature difference between the sample edges is denoted by $\Delta T$. A square-wave-modulated AC charge current $\mathbf{J}_c$ with amplitude $J_c$, frequency $f = 5$ Hz, and zero offset and an out-of-plane $\mathbf{H}$ with magnitude $H$ were applied along the $x$ and $z$ directions, respectively. When $\mathbf{J}_c$ and $\mathbf{H}$ are applied to the $Bi_{88}Sb_{12}$ sample, the Ettingshausen effect generates a heat current $\mathbf{J}_q$ in the $y$ direction, causing a temperature change at the $y$-axis sample edges (Fig. 1e). When $\nabla T$, $\mathbf{J}_c$, and $\mathbf{H}$ are co-applied, TTE heats or cools the entire sample. The resulting temperature modulation was measured using an infrared camera. The first harmonic component of the temperature modulation was extracted from the thermal images using the Fourier analysis and converted to lock-in amplitude $A$ and phase $\phi$ images to selectively detect the thermal response proportional to $J_c$, that is, the temperature modulation due to thermoelectric effects. The influence of the Peltier effect at the $x$-axis edges was minimized by capturing the center of the $x$-axis long sample (note that the Peltier effect generates $\mathbf{J}_q$ in the $x$ direction). According to equation (3), the sign of the TTE-induced temperature change should reverse by reversing $\mathbf{H}$; thus, we obtain the $H$-odd-dependent thermoelectric signals through $A_{odd} = |A(+H)\exp[-i\phi(+H)] - A(-H)\exp[-i\phi(-H)]|/2$ and $\phi_{odd} = -\arg[A(+H)\exp[-i\phi(+H)] - A(-H)\exp[-i\phi(-H)]]$ from thermal images captured under positive and negative magnetic fields to eliminate field-independent and $H$-even-dependent signals.

Figure 2d shows the images of steady-state $T$, $A_{odd}$, and $\phi_{odd}$ for the $Bi_{88}Sb_{12}$ sample at $J_c = 1$

A, $\Delta T = 40$ K, and $\mu_0 H = 200$ mT, where $\mu_0$ is the vacuum permeability. The steady-state $T$ image confirms that applied $\nabla T$ is uniform along the $y$ direction. The $A_{odd}$ image shows that a temperature change occurs around the edges of the sample. The $\phi_{odd}$ values of this thermal response differ by 180º between the right and left edges. The temperature change on the right (left) edge has almost the same phase as (opposite phase to) the input current and represents increasing (decreasing) $T$, indicating that the temperature change signals are due to the Ettingshausen effect. Under these conditions, the volumetric thermal response caused by TTE is expected to appear throughout the sample. However, the Ettingshausen signal at the sample edges obscures the TTE signals (Fig. 2d). To separate TTE from the Ettingshausen signal, we measured only the Ettingshausen effect in the absence of $\nabla T$ (at $\Delta T = 0$ K), while maintaining the average sample temperature at 320 K (Fig. 2e). The TTE signal can be extracted from the difference between the results of Fig. 2d and 2e, that is, by calculating $A_{diff} = |A_{odd}(\Delta T)\exp[-i\phi_{odd}(\Delta T)] - A_{odd}(\Delta T = 0$ K$)\exp[-i\phi_{odd}(\Delta T = 0$ K$)]|$ and $\phi_{diff} = -\arg[A_{odd}(\Delta T)\exp[-i\phi_{odd}(\Delta T)] - A_{odd}(\Delta T = 0$ K$)\exp[-i\phi_{odd}(\Delta T = 0$ K$)]]$ to eliminate the Ettingshausen background except for its $T$ dependence. Figure 2f shows the steady-state temperature difference $T_{diff} = T(\Delta T) - T(\Delta T = 0$ K$)$, $A_{diff}$, and $\phi_{diff}$ images at $J_c = 1$ A, $\Delta T = 40$ K, and $\mu_0 H = 200$ mT. Except for the regions near the sample edges, the sample exhibits clear volumetric temperature change, where $A_{diff}$ is almost uniform and $\phi_{diff}$ is constant around 90º. This $\phi_{diff}$ value is the same as the phase due to Joule heating[29]; therefore, the observed temperature change originates from volumetric heat release. Near the sample edges, the volumetric signal is obscured by a residual Ettingshausen signal due to its $T$ dependence. Nevertheless, this volumetric temperature change signal is well separated from the residual Ettingshausen signal around the center of the sample because the heat diffusion length of $Bi_{88}Sb_{12}$ at 5 Hz is $\sim 0.4$ mm, much smaller than the sample width (Methods).

To confirm the fundamental properties of the observed volumetric temperature change, we measured the $J_c$ and $\Delta T$ dependences of $A_{diff}$ and $\phi_{diff}$ systematically. Figure 3a presents the $A_{diff}$ and $\phi_{diff}$ images for the $Bi_{88}Sb_{12}$ sample for various values of $J_c$ at $\Delta T = 40$ K and $\mu_0 H = 200$ mT. As shown in the $y$-axis profiles of $A_{diff}$ and $\phi_{diff}$ in Fig. 3b, the thermal signal reaches its maximum value at the sample center. Hereafter, we discuss the detailed behavior using the averaged values of $A_{diff}$ and $\phi_{diff}$ taken within a 1 mm width at the center, unaffected by the edges. We found that $A_{diff}$ increased in proportion to $J_c$ and $\phi_{diff}$ remained unchanged with respect to $J_c$ (Fig. 3c). Figure 3d-f shows the $\Delta T$ dependence of $A_{diff}$ and $\phi_{diff}$ at $J_c = 1$ A and $\mu_0 H = 200$ mT. $A_{diff}$ is also proportional to $\Delta T$ and $\phi_{diff}$ is independent of $\Delta T$. These features of the volumetric thermal response satisfy equation (3) with respect to $J_c$ and $\Delta T$. Significantly, the volumetric temperature change disappears when $\mathbf{H} \parallel \mathbf{J}_c$ (Extended Data Fig. 1). This behavior is consistent with the features of TTE.

Figure 4a,b presents the $A_{diff}$ and $\phi_{diff}$ images and their line profiles along the $y$ direction for the same $Bi_{88}Sb_{12}$ sample at various $H$ values at $J_c = 1$ A and $\Delta T = 40$ K. We found that as $H$ increased, $A_{diff}$ peaked at 200 mT, decreased to 400 mT, and increased again for $\mu_0 H > 400$ mT (upper graph of Fig. 4c). $\phi_{diff}$ shifted from 90º to −90º at around 400 mT, indicating a reversal in the thermal response from heating to cooling (lower graph of Fig. 4c). This thermal response reversal occurs near the $H$ value where $\tau_{TTE}$ changes sign and cannot be explained by the behavior of $T(dS_{NE}/dT)$, which remains positive at all $H$ values (Fig. 2b).

We performed a numerical simulation of the thermoelectric transport for the $Bi_{88}Sb_{12}$ sample using the finite difference method to clarify the origin of the $H$ dependence of the volumetric temperature change in detail (Methods). The simulated $y$-axis profiles of $A_{diff}$ and $\phi_{diff}$ in Fig. 4d show the quantitatively estimated thermal responses of TTE based on equation (3) by substituting the measured $S_{NE}$ values and thermophysical properties of $Bi_{88}Sb_{12}$ in the transport model (Extended Data Fig. 2). The behaviors of the experimentally observed and calculated $A_{diff}$ and $\phi_{diff}$ values were similar (Fig. 4b,d). The magnitude of the simulated $A_{diff}$ was larger than that of the experimental one because we assumed that all heat due to TTE contributed to the change in the sample temperature by neglecting the heat loss to the sample holder in the simulation. The calculated $H$ dependence of the temperature modulation induced by $\tau_{TTE}$ in Fig. 4e indicates that $A_{diff}$ has two peaks at 200 and 700 mT and $\phi_{diff}$ changes from 90° to −90° around 400 mT, well consistent with the experimental results in Fig. 4c. This result implies that although the $H$ values at which the sign reversal occurs are slightly different between the observed lock-in thermography signals and $\tau_{TTE}$ estimated from measured $S_{NE}$, its difference can be quantitatively explained by the simulation considering the $T$ or position dependence of $\tau_{TTE}$ due to applied $\nabla T$. Therefore, all the observed behaviors of the volumetric thermal change relative to $J_c$, $\Delta T$, and $H$ align with equation (3), providing evidence of TTE observation. We revealed that the observable physical quantity of TTE is $\tau_{TTE}$, not $T(dS_{NE}/dT)$, which is a key distinction from the longitudinal Thomson effect.

To further investigate the behaviors of TTE, we divided the $A_{diff}$ and $\phi_{diff}$ signals into $T(dS_{NE}/dT)$ and $2S_{NE}$ components based on the simulation. Figure 5a,b shows the calculation results for the $T(dS_{NE}/dT)$ component. $\phi_{diff}$ remains 90° except for the $H$ region with tiny temperature dependence of $S_{NE}$, indicating that this component mainly induces the volumetric heat release due to positive $T(dS_{NE}/dT)$ of $Bi_{88}Sb_{12}$ (Fig. 2b). Conversely, the $2S_{NE}$ component induces the volumetric heat absorption, where $\phi_{diff}$ is always −90° (Fig. 5c,d). The magnitude of the thermal response due to $2S_{NE}$ increases with increasing $H$, consistent with the $H$ dependence of $|S_{NE}|$ in Fig. 2a. Thus, at low (high) $H$, the $T(dS_{NE}/dT)$ ($2S_{NE}$) component dominates the total TTE signal in $Bi_{88}Sb_{12}$, leading to the sign reversal of the volumetric temperature change.

The observation of TTE fills a missing piece in thermoelectrics, providing new insights into condensed matter physics and thermal management applications. As demonstrated in this study, the sign of the TTE-induced temperature change is controlled by the direction of the magnetic field and the magnitude of the temperature change is determined not only by the temperature derivative of $S_{NE}$ but also by the magnitude of $S_{NE}$. This feature distinguishes TTE from the longitudinal Thomson effect determined by the first Thomson relation. In our $Bi_{88}Sb_{12}$, the signs of the $T(dS_{NE}/dT)$ and $2S_{NE}$ components are opposite around room temperature, resulting in the sign reversal of the volumetric temperature change in its field dependence and the compensation of the thermoelectric performance of TTE. In other words, in materials where both components in $\tau_{TTE}$ have the same sign, the performance of TTE can be further improved. The observation of TTE in nonmagnetic materials suggests the existence of anomalous TTE in magnetic materials, which depends on the direction of spontaneous magnetization. Similar to TTE in nonmagnetic materials, anomalous TTE should become apparent in materials where both the temperature derivative and magnitude of the anomalous

Nernst coefficient are large. Observing anomalous TTE remains a task for future work; however, this phenomenon can be used to drive TTE-based thermal devices without applying an external magnetic field. The experimental methods established in this study will be useful for observing such unobserved higher-order thermoelectric phenomena.

## Methods

### Measurement procedures

We prepared a Bi88Sb12 sample with lengths of 13.0, 3.1, and 0.3 mm along the $x$, $y$, and $z$ directions, respectively, to measure TTE by the lock-in thermography method. The side edges of the Bi88Sb12 sample were fixed to two plates using thermal conductive silicone adhesive, where the contact area between the sample and plate was $13.0 \times 0.5$ mm$^2$ (Fig. 2c). The plates were made of anodized aluminum and were electrically insulated from the sample. A ceramic heater and Peltier module were attached to each plate to control the magnitude of the applied temperature gradient and the average temperature of the sample. Cu wires were electrically connected to the $3.1 \times 0.3$ mm$^2$ surfaces of the Bi88Sb12 slab using indium. A magnetic field was applied to the sample in the $z$ direction by an electrical magnet, with a maximum magnetic field of 700 mT. An insulating black ink with an infrared emissivity $> 0.94$ was coated on the sample surface to increase the emissivity and ensure its uniformity. We performed lock-in thermography measurements after the temperature gradient and average temperature of the sample reached a steady state. The measurement time for the $A$ and $\phi$ images for each condition was 10 min.

### Formulation of Nernst and Ettingshausen effects

From Ohm's law, Fourier's law, and Onsager reciprocal relations, the linear response relations of the charge current density $\mathbf{j}_c$ and heat current density $\mathbf{j}_q$ can be written as[20]

$$\begin{pmatrix} \mathbf{j}_c \\ \mathbf{j}_q \end{pmatrix} = \begin{pmatrix} \sigma & \sigma S T \\ \sigma S T & \kappa T \end{pmatrix} \begin{pmatrix} \mathbf{E} \\ -\nabla T / T \end{pmatrix} \qquad (5)$$

where $\mathbf{E}$, $\sigma$, $S$, and $\kappa$ denote the electric field, electrical conductivity, Seebeck coefficient, and thermal conductivity tensors, respectively. The matrix formula commonly used in the field of thermoelectrics is obtained by replacing $\mathbf{j}_c$ and $\mathbf{E}$ in equation (5) as follows:

$$\begin{pmatrix} \mathbf{E} \\ \mathbf{j}_q \end{pmatrix} = \begin{pmatrix} \rho & S \\ ST & -\kappa' \end{pmatrix} \begin{pmatrix} \mathbf{j}_c \\ \nabla T \end{pmatrix} \qquad (6)$$

where $\rho \ (= \sigma^{-1})$ is the electrical resistivity and $\kappa' \equiv \kappa - \sigma S^2 T$. When a magnetic field is applied in the $z$ direction, equation (6) in the $x$-$y$ plane is described as

$$\begin{pmatrix} E_x \\ E_y \\ j_{q,x} \\ j_{q,y} \end{pmatrix} = \begin{pmatrix} \rho_{xx} & \rho_{xy} & S_{xx} & S_{xy} \\ -\rho_{xy} & \rho_{xx} & -S_{xy} & S_{xx} \\ S_{xx}T & S_{xy}T & -\kappa'_{xx} & -\kappa'_{xy} \\ -S_{xy}T & S_{xx}T & \kappa'_{xy} & -\kappa'_{xx} \end{pmatrix} \begin{pmatrix} j_{c,x} \\ j_{c,y} \\ \nabla_x T \\ \nabla_y T \end{pmatrix} \qquad (7)$$

in an isotropic material, where $\rho_{xx} = \rho_{yy}$, $S_{xx} = S_{yy}$, $\kappa'_{xx} = \kappa'_{yy}$, $\rho_{xy} = -\rho_{yx}$, $S_{xy} = -S_{yx}$, and $\kappa'_{xy} = -\kappa'_{yx}$.

First, we consider the Nernst-effect-induced electric field $E_y$ when a temperature gradient $\nabla_x T$ and magnetic field $H_z$ are applied in the $x$ and $z$ directions, respectively. In the Nernst measurement, the $x$ and $y$ directions are typically under open-circuit conditions ($j_{c,x} = j_{c,y} = 0$). Here, assuming the isothermal condition in the $y$ direction ($\nabla_y T = 0$), the measured thermopower is obtained from equation (7) as

$$\frac{E_{y,\mathrm{i}}}{-\nabla_x T} = S_{xy} \qquad (8)$$

where the subscript i denotes the isothermal condition. Thus, the pure Nernst coefficient can be obtained from Nernst measurements under the isothermal condition in the transverse direction. In contrast, assuming the adiabatic condition in the $y$ direction ($j_{q,y} = 0$), the thermopower is calculated as

$$\frac{E_{y,\mathrm{a}}}{-\nabla_x T} = S_{xy} - \theta_{\mathrm{th}} S_{xx} \equiv S^*_{xy} \qquad (9)$$

where the subscript a denotes the adiabatic condition and $\theta_{\mathrm{th}}$ ($\equiv \kappa'_{xy}/\kappa'_{xx}$) is the thermal Hall angle. Therefore, the observed thermopower includes not only the contribution from the Nernst effect but also that from the Seebeck and thermal Hall effects, making it difficult to extract the pure $S_{xy}$ contribution.

Next, we consider the temperature gradient $\nabla_y T$ induced by the Ettingshausen effect when a charge current density $j_{c,x}$ and magnetic field $H_z$ are applied in the $x$ and $z$ directions, respectively. In the Ettingshausen measurement, the $y$ direction is typically under an open-circuit condition ($j_{c,y} = 0$). Here, assuming the isothermal condition in the $x$ direction ($\nabla_x T = 0$), the measured charge-to-heat current conversion ratio is obtained from equation (7) as

$$\frac{-\kappa'_{xx}(\nabla_y T)_{\mathrm{i}}}{j_{c,x}} = S_{xy} T \qquad (10)$$

Thus, the pure Nernst coefficient can be obtained from Ettingshausen measurement under the isothermal condition in the longitudinal direction. In contrast, assuming the adiabatic condition in the $x$ direction ($j_{q,x} = 0$), the charge-to-heat current conversion ratio is calculated as

$$\frac{-\kappa'_{xx}(\nabla_y T)_{\mathrm{a}}}{j_{c,x}} = (S_{xy} - \theta_{\mathrm{th}} S_{xx})T = S^*_{xy} T \qquad (11)$$

where $\kappa'_{xy} << \kappa'_{xx}$ is assumed. Similar to the Nernst measurements, under the adiabatic condition, the observed temperature change signal includes not only the contribution from the Ettingshausen effect but also that from the Peltier and thermal Hall effects considering the Onsager reciprocal relations.

Whether isothermal or adiabatic conditions should be applied depends on the aspect ratio of the sample in the $x$-$y$ plane[30]. In a rectangular sample, when the $x$-axis length is much larger than the $y$-axis length, the sample is nearly isothermal in the $x$ direction and adiabatic in the $y$ direction. As described above, the Nernst and Ettingshausen measurements are affected by the thermal boundary conditions along the electric field, that is, the $y$ and $x$ directions, respectively. Consequently, the Ettingshausen measurements are better suited for estimating $S_{xy}$ than the Nernst measurements for the aspect ratio of our sample. Thus, the $T$ dependence of the Nernst coefficient in Fig. 2b was estimated through the Ettingshausen measurements using the lock-in thermography method, in which the influence of the Peltier and thermal Hall effects was suppressed.

**Formulation of longitudinal Thomson effect**

We consider the heat production rate per unit volume $\dot{q}$ generated when a charge current density $j_{c,x}$ and temperature gradient $\nabla_x T$ are applied in the $x$ direction. $\dot{q}$ comprises the Joule heating and the divergence of the heat current and is obtained using equation (7) as follows:

$$
\begin{aligned}
\dot{q} &= E_x j_{c,x} - \operatorname{div} j_{q,x} \\
&= (\rho_{xx} j_{c,x} + S_{xx} \nabla_x T) j_{c,x} - \nabla_x (S_{xx} T j_{c,x} - \kappa'_{xx} \nabla_x T) \\
&= \rho_{xx} j_{c,x}^2 + \nabla_x (\kappa'_{xx} \nabla_x T) + \left( S_{xx} - \frac{dS_{xx} T}{dT} \right) j_{c,x} \nabla_x T \\
&= \rho_{xx} j_{c,x}^2 + \nabla_x (\kappa'_{xx} \nabla_x T) + \left( S_{xx} - S_{xx} - T \frac{dS_{xx}}{dT} \right) j_{c,x} \nabla_x T \\
&= \rho_{xx} j_{c,x}^2 + \nabla_x (\kappa'_{xx} \nabla_x T) - T \frac{dS_{xx}}{dT} j_{c,x} \nabla_x T
\end{aligned}
\tag{12}
$$

The last term of the last line in equation (12) is the heat generated by the concerted action of $j_{c,x}$ and $\nabla_x T$, that is, the longitudinal Thomson effect (equations (1) and (2)). The last term of the fourth line in equation (12) shows that the $S_{xx}$ components contributing to the longitudinal Thomson effect generated by the Joule heating and heat-current divergence terms have different signs and eliminate each other. Hence, the observable physical quantity of the longitudinal Thomson effect is determined only by the temperature derivative of the Seebeck effect.

**Formulation of transverse Thomson effect**

We consider the heat production rate per unit volume $\dot{q}$ generated when a charge current density $j_{c,x}$, temperature gradient $\nabla_y T$, and magnetic field $H_z$ are applied in the $x$, $y$, and $z$ directions, respectively. The $y$ direction is assumed to be under the open-circuit condition ($j_{c,y} = 0$). We solve $\dot{q}$ for two thermal boundary conditions in the $x$ direction. First, $\dot{q}_i$ for the isothermal condition in the $x$ direction ($\nabla_x T = 0$), is obtained using equation (7) as follows:

$$
\begin{aligned}
\dot{q}_i &= E_x j_{c,x} - \operatorname{div} j_{q,y} \\
&= (\rho_{xx} j_{c,x} + S_{xy} \nabla_y T) j_{c,x} - \nabla_y (-S_{xy} T j_{c,x} - \kappa'_{xx} \nabla_y T) \\
&= \rho_{xx} j_{c,x}^2 + \nabla_y (\kappa'_{xx} \nabla_y T) + \left( S_{xy} + \frac{dS_{xy} T}{dT} \right) j_{c,x} \nabla_y T \\
&= \rho_{xx} j_{c,x}^2 + \nabla_y (\kappa'_{xx} \nabla_y T) + \left( S_{xy} + S_{xy} + T \frac{dS_{xy}}{dT} \right) j_{c,x} \nabla_y T \\
&= \rho_{xx} j_{c,x}^2 + \nabla_y (\kappa'_{xx} \nabla_y T) + \left( T \frac{dS_{xy}}{dT} + 2 S_{xy} \right) j_{c,x} \nabla_y T
\end{aligned}
\tag{13}
$$

When the magnetic field is reversed, the first and second terms of the last line in equation (13) do not change their sign and only the last term reverses its sign. This is because the first and second terms are determined only by the diagonal components of the transport tensors, whereas the last term is determined by the off-diagonal component. The last term of the last line indicates the volumetric heat release or absorption due to TTE, which is generated by the concerted action of $j_{c,x}$ and $\nabla_y T$. The proportionality factor of the $H$-odd-dependent term is defined as $\tau_{\text{TTE}}$ to be

$$\tau_{\text{TTE}} \equiv T\frac{dS_{xy}}{dT} + 2S_{xy} \tag{14}$$

Therefore, experimentally measured TTE includes not only the temperature derivative of the Nernst coefficient but also its magnitude. This feature differs from the longitudinal Thomson coefficient, which is determined only by the temperature derivative of the Seebeck coefficient (equation (2)). The reason for this difference is that the $S_{xy}$ components of TTE generated by the Joule heating and heat-current divergence terms have the same sign and enhance each other (see the last term of the fourth line in equation (13)).

Next, we derive $\dot{q}_a$ for the adiabatic condition in the $x$ direction ($j_{q,x} = 0$). From equation (7), $\nabla_x T$ is generated by the heat current due to the Peltier and thermal Hall effects:

$$\nabla_x T = \frac{S_{xx}Tj_{c,x} - \kappa'_{xy}\nabla_y T}{\kappa'_{xx}} \tag{15}$$

$\dot{q}_a$ with finite $\nabla_x T$ in equation (15) can be obtained as

$$
\begin{aligned}
\dot{q}_a &= E_x j_{c,x} - \operatorname{div} j_{q,y} \\
&= (\rho_{xx}j_{c,x} + S_{xx}\nabla_x T + S_{xy}\nabla_y T)j_{c,x} - \nabla_y(-S_{xy}Tj_{c,x} + \kappa'_{xy}\nabla_x T - \kappa'_{xx}\nabla_y T) \\
&= \rho_{xx}j_{c,x}^2\left(1 + \frac{S_{xx}^2 T}{\rho_{xx}\kappa'_{xx}}\right) + \nabla_y\left(\frac{\kappa'^2_{xx} + \kappa'^2_{xy}}{\kappa'_{xx}}\nabla_y T\right) \\
&\quad + \left[(S_{xy} - \theta_{\text{th}}S_{xx}) + \frac{d}{dT}\big((S_{xy} - \theta_{\text{th}}S_{xx})T\big)\right]j_{c,x}\nabla_y T \\
&= \rho_{xx}j_{c,x}^2\left(1 + \frac{S_{xx}^2 T}{\rho_{xx}\kappa'_{xx}}\right) + \nabla_y\left(\frac{\kappa'^2_{xx} + \kappa'^2_{xy}}{\kappa'_{xx}}\nabla_y T\right) \\
&\quad + \left[T\frac{d(S_{xy} - \theta_{\text{th}}S_{xx})}{dT} + 2(S_{xy} - \theta_{\text{th}}S_{xx})\right]j_{c,x}\nabla_y T
\end{aligned}
\tag{16}
$$

The last term of the last line in equation (16), due to TTE caused by $j_{c,x}$ and $\nabla_y T$, includes the contribution of the Seebeck and thermal Hall effects and differs from that in the isothermal condition. Therefore, the transverse Thomson coefficient $\tau^*_{\text{TTE}}$ in the adiabatic condition is

$$\tau^*_{\text{TTE}} = T\frac{dS^*_{xy}}{dT} + 2S^*_{xy} \tag{17}$$

indicating that $\tau^*_{\text{TTE}}$ is determined by $S^*_{xy}$ and affected by the thermal boundary conditions.

In this study, we used a rectangular sample with a longer $x$-axis length than a $y$-axis length for measuring TTE to reduce $\nabla_x T$ occurrence. Thus, we have discussed TTE assuming the isothermal condition.

**Numerical simulation for TTE**

The TTE experiment was simulated using a numerical model for the $Bi_{88}Sb_{12}$ sample based on the finite-difference method. The simulation was performed for the $y$ direction using the following unsteady one-dimensional heat equation:

$$\rho_{d} C \frac{\partial T}{\partial t} = \kappa \frac{\partial^2 T}{\partial y^2} + \dot{q} \tag{18}$$

where $\rho_d$ and $C$ are the density and specific heat, respectively. The heat source $\dot{q}$ in equation (18) includes the contributions from the Ettingshausen effect at the edges of the sample and TTE occurring throughout the sample as follows:

$$\begin{cases} \dot{q}(y = 0 \text{ mm}) = \dfrac{S_{NE} T j_{c,x}}{\Delta y} & \\[2mm] \dot{q}(y = 3.1 \text{ mm}) = -\dfrac{S_{NE} T j_{c,x}}{\Delta y} & \text{due to the Ettingshausen effect} \\[2mm] \dot{q}(y) = \left( T \dfrac{dS_{NE}}{dT} + 2S_{NE} \right) j_{c,x} \nabla_y T & \text{due to TTE} \end{cases} \tag{19}$$

where $\Delta y$ is the length of one node and the sample is divided into more than 100 nodes in the $y$ direction. The equations above were calculated under the adiabatic condition: $j_{q,y}(y = 0 \text{ mm}) = j_{q,y}(y = 3.1 \text{ mm}) = 0$. The $y$-axis profiles of $A_{odd}$ and $\phi_{odd}$ were derived using Fourier analysis from the temperature response of each node after the calculations converged when a square-wave-modulated AC charge current $j_{c,x}$ was applied. The $j_{c,x}$ value was assumed to be constant at each node because $\sigma$ of $Bi_{88}Sb_{12}$ changes only ±3 % with respect to $T$ for all $H$ and its $H$ dependence has no effect on TTE (Extended Data Fig. 2). We performed this calculation with and without applying $\nabla_y T$. The $A_{diff}$ and $\phi_{diff}$ values were obtained by determining the difference between the results. We used the measured $S_{NE}$ (Fig. 2b) and thermophysical properties ($\rho_d$ = 9260 kg m$^{-3}$, $C$ = 135 J kg$^{-1}$ K$^{-1}$, and $\kappa$ = 3.9 W m$^{-1}$ K$^{-1}$ presented in Extended Data Fig. 2) of $Bi_{88}Sb_{12}$ for the simulation. As $C$ and $\kappa$ do not depend on $T$ and $\kappa$ varies only by ±5 % under $H$, the $C$ and $\kappa$ values were averaged and regarded as the constant parameters.

**Data availability**

The data that support the findings of this study are available from the corresponding author upon reasonable request.

**Acknowledgements**  The authors thank R. Iguchi and F. Ando for valuable discussions and M. Isomura and K. Suzuki for technical supports. This work was supported by JST ERATO "Magnetic Thermal Management Materials" (JPMJER2201) from JST, Japan; Grant-in-Aid for Scientific Research (B) (19H02585) and Grant-in-Aid for Scientific Research (S) (22H04965) from JSPS, Japan; and NEC Corporation. A.T. was supported by Grant-in-Aid for JSPS Fellows (23KJ1122) from JSPS, Japan.

**Author contributions**  K.U. planned the study; K.U. and H.N. supervised the study; A.T. and K.U. designed the experiments, prepared the samples, verified the formulations for transverse thermoelectric effects, developed the explanation of the experiments, and prepared the manuscript; A.T. performed the experiments for TTE with support from T.H. and performed the numerical simulation; A.A., S.J.P., and A.T. measured the thermophysical properties of the sample. All the authors discussed the results and commented on the manuscript.

**Competing interests**  The authors declare no competing interests.

| Input<br>Configuration | Heat current | Charge current | Heat and charge currents |
|---|---|---|---|
| Longitudinal | **a** Seebeck effect (1821) 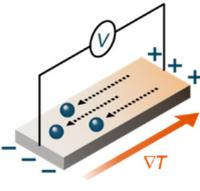 | **b** Peltier effect (1834) 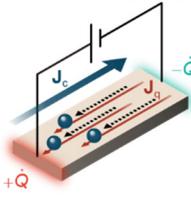 | **c** Thomson effect (1856) 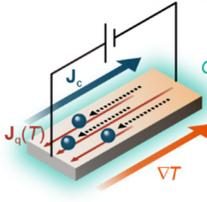 |
| Transverse | **d** Nernst effect (1886) 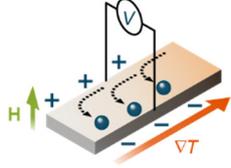 | **e** Ettingshausen effect (1887) 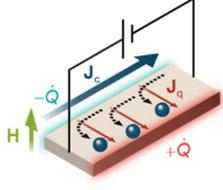 | **f** Transverse Thomson effect (2025) 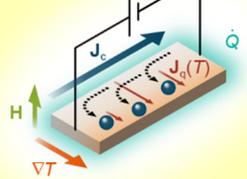 |

**Fig. 1: Thermoelectric effects. a,** Schematic of the Seebeck effect. When a temperature gradient $\nabla T$ is applied to a conductor, an electric voltage $V$ is generated along the $\nabla T$ direction. **b,** Schematic of the Peltier effect. When a charge current $\mathbf{J}_c$ is applied to a conductor, a heat current $\mathbf{J}_q$ is generated along the $\mathbf{J}_c$ direction and induces a heat release or absorption $\pm \dot{Q}$ at the edges of the conductor, that is, the interfaces with different materials. **c,** Schematic of the Thomson effect. When $\mathbf{J}_c$ and $\nabla T$ are applied to a conductor in parallel directions, volumetric $\dot{Q}$ is generated throughout the conductor. **d,** Schematic of the Nernst effect. When $\nabla T$ is applied to a conductor, $V$ is generated in the direction of the cross product of $\nabla T$ and a magnetic field $\mathbf{H}$. **e,** Schematic of the Ettingshausen effect. When $\mathbf{J}_c$ is applied to a conductor, $\mathbf{J}_q$ is generated in the direction of the cross product of $\mathbf{J}_c$ and $\mathbf{H}$ and induces $\pm \dot{Q}$ at the side edges of the conductor. **f,** Schematic of the transverse Thomson effect. When $\mathbf{J}_c$, $\nabla T$, and $\mathbf{H}$ are applied to a conductor in perpendicular directions, volumetric $\dot{Q}$ is generated throughout the conductor. The blue spheres in the schematics represent electron transport. In semiconductors and semimetals, hole transport also contributes to these thermoelectric effects in the same manner.

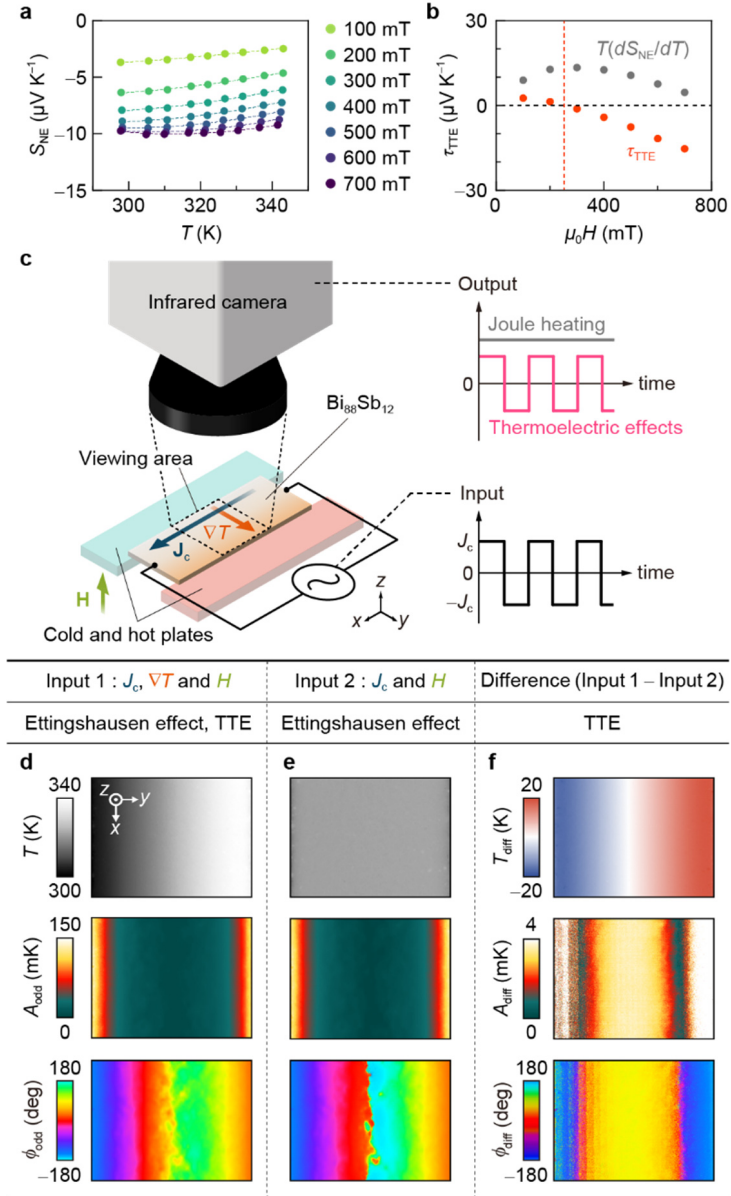

**Fig. 2: Transverse thermoelectric conversion properties and lock-in thermography measurements. a,** $T$ dependence of the Nernst coefficient $S_{NE}$ of $Bi_{88}Sb_{12}$ for various values of the magnetic field magnitude $H$. **b,** $H$ dependence of the transverse Thomson coefficient $\tau_{TTE}$ (equation (4)) and its component $T(dS_{NE}/dT)$ at 320 K calculated using the data in **a**. **c,** Schematic of the set-up for the lock-in thermography measurement of TTE. To excite TTE, $\mathbf{J}_c$ (with magnitude $J_c$), $\nabla T$ (with the temperature difference between the sample edges $\Delta T$), and $\mathbf{H}$ were applied along the $x$, $y$, and $z$ directions, respectively. **d,** Steady-state temperature $T$, $H$-odd-dependent lock-in amplitude $A_{odd}$, and phase $\phi_{odd}$ images for $Bi_{88}Sb_{12}$ at $J_c = 1$ A, $\Delta T = 40$ K, and $\mu_0 H = 200$ mT. **e,** Steady-state $T$, $A_{odd}$, and $\phi_{odd}$ images at $J_c = 1$ A, $\Delta T = 0$ K, and $\mu_0 H = 200$ mT. **f,** Steady-state temperature $T_{diff}$, amplitude $A_{diff}$, and phase $\phi_{diff}$ images obtained by calculating the difference between the results of **d** and **e**.

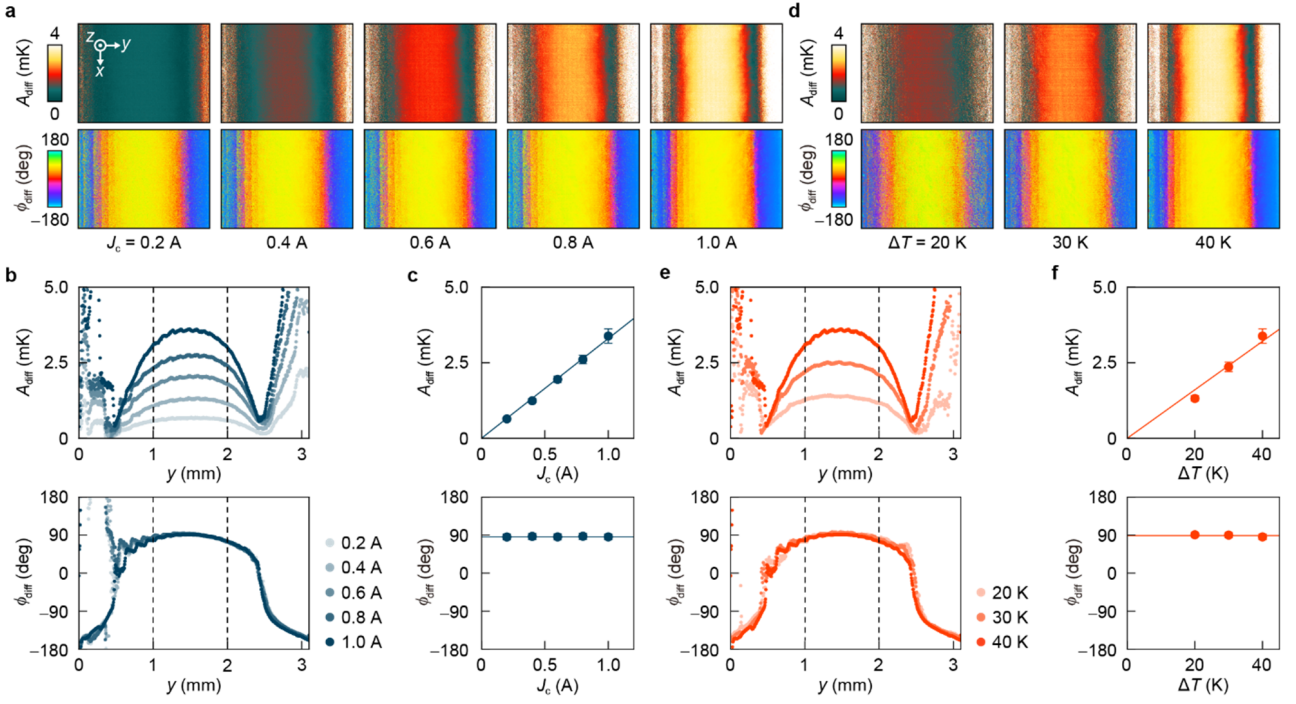

**Fig. 3: Charge current and temperature gradient dependences of volumetric temperature change. a,** $A_{\text{diff}}$ and $\phi_{\text{diff}}$ images for the Bi$_{88}$Sb$_{12}$ sample for various values of $J_{\text{c}}$ at $\Delta T = 40$ K and $\mu_0 H = 200$ mT. **b,** $y$-axis profiles of $A_{\text{diff}}$ and $\phi_{\text{diff}}$ for various values of $J_{\text{c}}$. **c,** $J_{\text{c}}$ dependence of $A_{\text{diff}}$ and $\phi_{\text{diff}}$ at $\Delta T = 40$ K and $\mu_0 H = 200$ mT. **d,** $A_{\text{diff}}$ and $\phi_{\text{diff}}$ images for various values of $\Delta T$ at $J_{\text{c}} = 1$ A and $\mu_0 H = 200$ mT. **e,** $y$-axis profiles of $A_{\text{diff}}$ and $\phi_{\text{diff}}$ for various values of $\Delta T$. **f,** $\Delta T$ dependence of $A_{\text{diff}}$ and $\phi_{\text{diff}}$ at $J_{\text{c}} = 1$ A and $\mu_0 H = 200$ mT. The $y$-axis profiles in **b** and **e** are obtained by averaging the raw profiles in the entire regions of the images in **a** and **d**. The data in **c** and **f** are obtained by averaging the $A_{\text{diff}}$ and $\phi_{\text{diff}}$ values between the dotted lines in **b** and **e**, respectively. The error bars represent the standard deviation of the data.

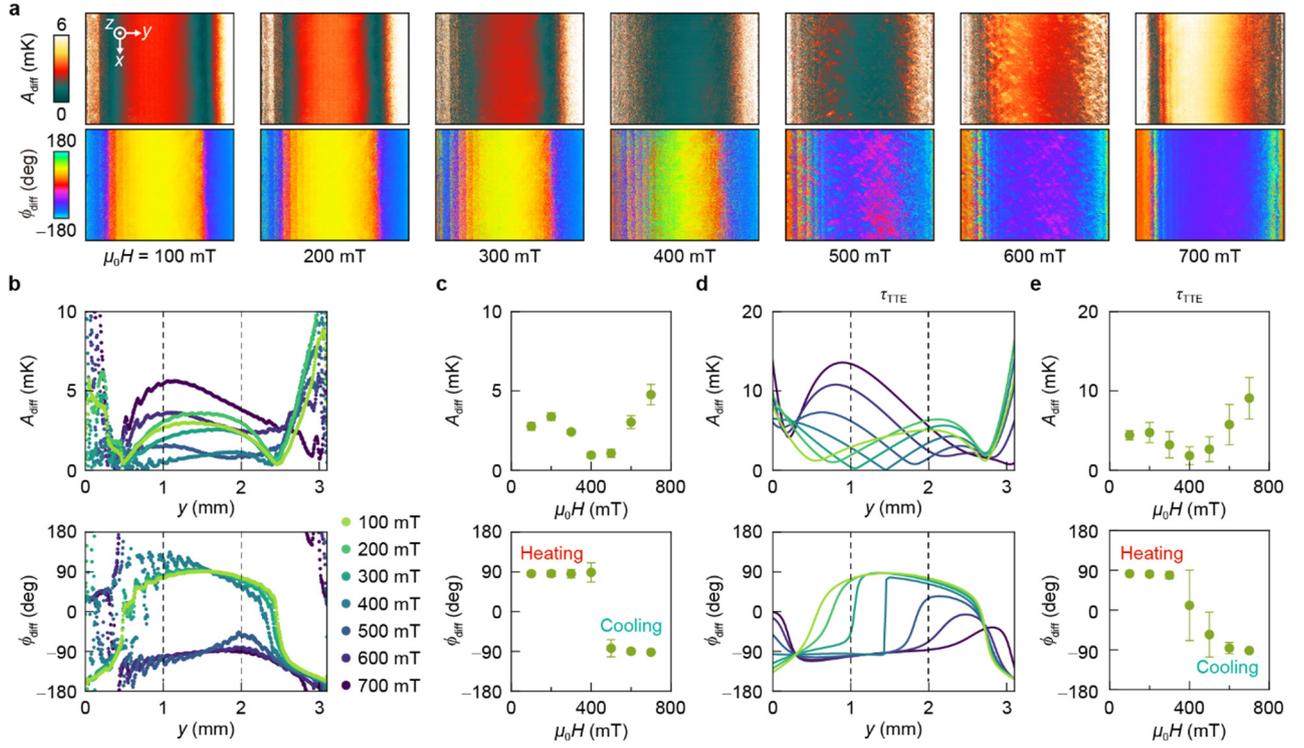

**Fig. 4: Magnetic field dependence of volumetric temperature change. a,** $A_{\text{diff}}$ and $\phi_{\text{diff}}$ images for the Bi$_{88}$Sb$_{12}$ sample for various $H$ values at $J_c$ = 1 A and $\Delta T$ = 40 K. **b,** Observed $y$-axis profiles of $A_{\text{diff}}$ and $\phi_{\text{diff}}$ for various $H$ values. **c,** Observed $H$ dependence of $A_{\text{diff}}$ and $\phi_{\text{diff}}$ at $J_c$ = 1 A and $\Delta T$ = 40 K. **d,** Simulated $y$-axis profiles of $A_{\text{diff}}$ and $\phi_{\text{diff}}$ for various $H$ values at $J_c$ = 1 A and $\Delta T$ = 40 K, obtained by substituting the experimentally determined thermoelectric and thermophysical properties of Bi$_{88}$Sb$_{12}$ into the finite difference model. **e,** Simulated $H$ dependence $A_{\text{diff}}$ and $\phi_{\text{diff}}$ at $J_c$ = 1 A and $\Delta T$ = 40 K. The $y$-axis profiles in **b** are obtained by averaging the raw profiles in the entire regions of the images in **a**. The data in **c** and **e** are obtained by averaging the $A_{\text{diff}}$ and $\phi_{\text{diff}}$ values between the dotted lines in **b** and **d**, respectively. The error bars represent the standard deviation of the data.

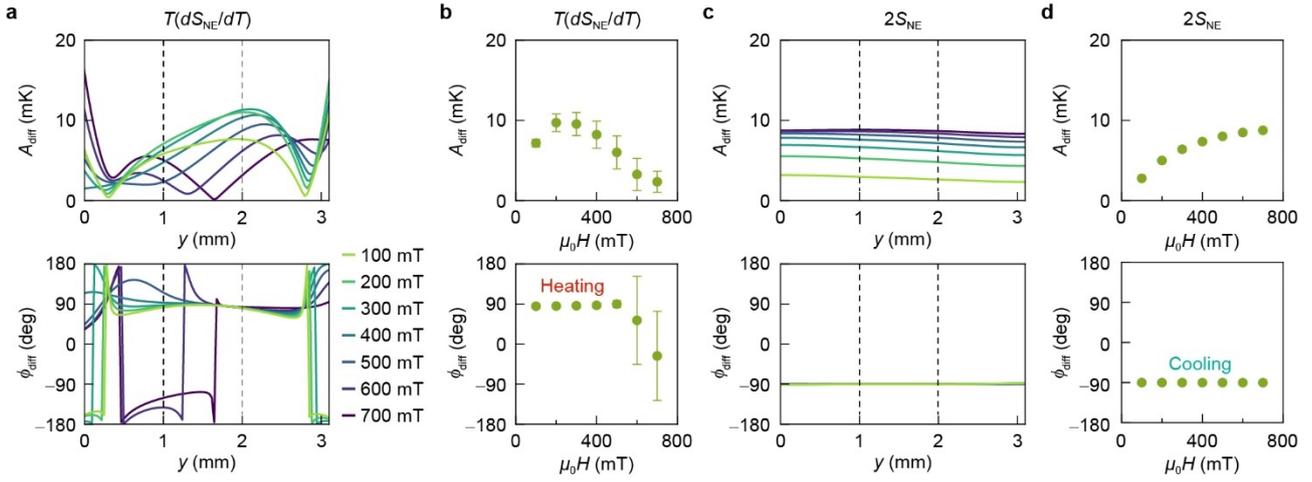

**Fig. 5: Verification of components of transverse Thomson coefficient. a,** Simulated $y$-axis profiles of $A_{\text{diff}}$ and $\phi_{\text{diff}}$ due to the $T(dS_{\text{NE}}/dT)$ component for the Bi$_{88}$Sb$_{12}$ sample for various values of $H$ at $J_{\text{c}} = 1$ A and $\Delta T = 40$ K. The data were calculated using the thermoelectric and thermophysical properties of Bi$_{88}$Sb$_{12}$ and the thermoelectric transport model based on the finite-difference method. **b,** Simulated $H$ dependence $A_{\text{diff}}$ and $\phi_{\text{diff}}$ due to the $T(dS_{\text{NE}}/dT)$ component at $J_{\text{c}} = 1$ A and $\Delta T = 40$ K. **c,** Simulated $y$-axis profiles of $A_{\text{diff}}$ and $\phi_{\text{diff}}$ due to the $2S_{\text{NE}}$ component for various $H$ values at $J_{\text{c}} = 1$ A and $\Delta T = 40$ K. **d,** Simulated $H$ dependence $A_{\text{diff}}$ and $\phi_{\text{diff}}$ due to the $2S_{\text{NE}}$ component at $J_{\text{c}} = 1$ A and $\Delta T = 40$ K. The data in **b** and **d** are obtained by averaging the $A_{\text{diff}}$ and $\phi_{\text{diff}}$ values between the dotted lines in **a** and **c**, respectively. The error bars represent the standard deviation of the data.

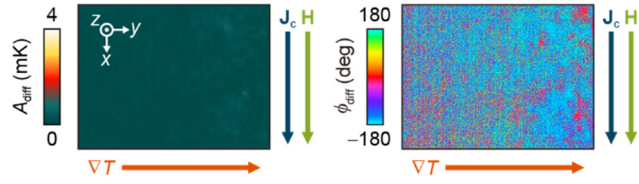

**Extended Data Fig. 1: Magnetic field direction dependence.** $A_{diff}$ and $\phi_{diff}$ images for the Bi$_{88}$Sb$_{12}$ sample at $J_c =$ 1 A, $\Delta T = 40$ K, and $\mu_0 H = 200$ mT for **H** ∥ **J**$_c$.

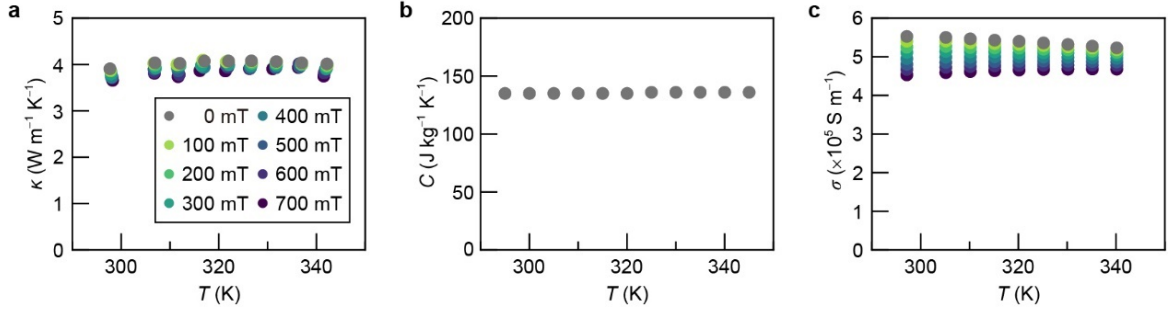

**Extended Data Fig. 2: Physical properties of Bi$_{88}$Sb$_{12}$. a,** $T$ dependence of the thermal conductivity $\kappa$ of Bi$_{88}$Sb$_{12}$ for various $H$ values. $\kappa$ was measured based on the method of ref.[31] using an electromagnet and a sample holder with $T$ control modules. **b,** $T$ dependence of the specific heat $C$ of Bi$_{88}$Sb$_{12}$ at zero field, measured using a differential scanning calorimeter. **c,** $T$ dependence of the electrical conductivity $\sigma$ of Bi$_{88}$Sb$_{12}$ for various $H$ values, measured using the four-probe method.